\documentstyle[12pt,aaspp4,flushrt]{article}
\def\etal{{\it et~al.~}}

\def\cf{{\it c.f.~}}
\def\HST{{\it Hubble Space Telescope}}

\def\ii{{\small II~}}

\def\arcsec{^{\prime\prime}}

\def\kpc{\mbox{kpc}}
\begin{document}

\title{Are Hubble Deep Field Galaxy Counts Whole Numbers?}

\author{Wesley N. Colley, \footnote{Supported by the
Fannie and John Hertz Foundation, Livermore, CA 94551-5032}
James E. Rhoads, Jeremiah P. Ostriker}
\affil{Princeton University Department of Astrophysical Sciences, 
Princeton, NJ 08544}
\author{and David N. Spergel
\footnote{on sabbatical from Princeton University Department of Astrophysical
Sciences}}
\affil{Department of Astronomy, University of Maryland, College Park, MD 20742}
\begin{center}
Email: wes,rhoads,jpo@astro.princeton.edu; dns@astro.umd.edu
\end{center}

\begin{abstract}

The Hubble Deep Field\footnote{Based on observations with the NASA/ESA \HST,
obtained at the Space Telescope Science Institute, which is operated by AURA,
under NASA contract NAS 5-26555} (HDF) offers the best view to date of the
optical sky at faint magnitudes and small angular scales.  Early reports
suggested that faint source counts continue to rise to the completeness limit
of the data, implying a very large number of galaxies.  In this {\it letter,}
we use the two-point angular correlation function and number-magnitude relation
of sources within the HDF in order to assess their nature.  We find that the
correlation peaks between $0.25\arcsec$ and $0.4\arcsec$ with amplitude of 2 or
greater, and much more for the smallest objects.  This angular scale
corresponds to physical scales of order $1 \kpc$ for redshifts $z \ga 1$.  The
correlation must therefore derive from objects with subgalaxian separations.
At faint magnitudes, the counts satisfy the relation $\mbox{Number} \propto
1/\mbox{flux}$, expected for images which are subdivisions of larger ones.

Several explanations for these observed correlations are possible, but a
conservative explanation can suffice to produce our results.  Since high
redshift space ($z \ga 0.5$) dominates the volume of the sample, observational
redshift effects are important.  Rest-frame ultraviolet radiation appears in
the HDF's visible and near-UV bands, and surface brightness dimming enhances
the relative brightness of unresolved objects versus resolved objects.  Both
work to increase the prominence of compact star-forming regions over diffuse
stellar populations.  Thus, a ``normal'' gas-rich galaxy at high redshift can
appear clumpy and asymmetric in the visible bands.  For sufficiently faint and
distant objects, the compact star-forming regions in such galaxies peak above
undetectable diffuse stellar backgrounds.  Our results do not exclude
asymmetric formation or fragmentation scenarios.

\end{abstract}


\section{Introduction}

The Hubble Deep Field (Williams \etal 1996) affords us an unprecedented view of
the optical sky at small angular scales and faint flux levels.  It thus allows
us to study faint (and presumably high redshift) galaxies without complicating
field crowding effects caused by comparatively poor seeing in ground-based
faint galaxy studies (\cf Tyson 1995).

Preliminary results (Giavalisco \etal 1996) show that source counts in the HDF
continue to rise as a power law to the completeness limit of the data.  Such an
effect may be due to ever larger numbers of galaxies at fainter flux levels.
However, it may also be due to the increasingly clumpy appearance of galaxies
at high redshift, which can confuse source detection algorithms into counting
parts of each physically distinct galaxy as several faint sources.

To test this possibility, we consider how redshift effects can lead to
over-counting of whole sources in a deep field like the HDF.  $K$-correction
and surface-brightness dimming tend to enhance the relative prominence UV
bright and compact objects, such as active star-forming regions (O'Connell \&
Marcum 1996).  If the enhancement is sufficient, several star-forming regions
occurring in a single galaxy will produce the appearance of several small
sources separated by an angular scale comparable to the size of a normal
galaxy.

In the later sections of the paper, we use two different statistics to test the
extent to which HDF source counts reflect the subdivision of galaxies.  These
tests exploit the weak dependence of angular size on redshift $z$ at $z \ga 1$.
Since galaxies of present day size ($10\kpc$) remain resolved at all redshifts
in the HDF (Peebles 1993), we can compare the physical separations and sizes of
objects in the HDF to those of galaxies in the low-redshift universe.

In section~4, we discuss the two-point angular correlation function $w(\theta)$
of HDF sources.  Considerable correlations may be expected for physical scales
$\la 10 \kpc$ if many galaxies in the field break up into multiple giant H\ii
regions in the source catalogs.

In section~5, we present number-magnitude relations derived from our source
catalogs, which show a smooth increase to the completeness limit, with a
flatter faint-end slope than in deep ground-based images, and a rough relation
$N \propto 1/\mbox{flux}$, consistent with the hypothesis that many of the
faintest images are parts of larger objects.

\section{Redshift effects and deep counts}

Two redshift effects play an important role in the appearance of very deep
fields.  First, the redshift moves the ultraviolet rest-frame light into the
optical, so that rest-frame UV-bright objects will be selected over optically
bright objects.  Schade \etal (1995) have found that in up to one-third
galaxies with $0.5<z<1.2$, compact blue components dominate the blue light.  In
such galaxies, active star-formation is occurring.  Abraham \etal (1996) and
Clements and Couch (1996) have found the trend toward increasingly blue
asymmetric objects to be even more pronounced in the HDF, where yet higher
redshifts ($z \la 4$) bring ever harder rest-frame UV emission into optical
bands.  Moreover, many spectroscopically confirmed high redshift objects in the
HDF display noticeable asymmetry and multiple structure.  A nice collection of
such objects can be found in figure~2 of Steidel \etal (1996).

Second, compact high redshift objects can appear more prominently than diffuse
objects if their angular size is smaller than the point-spread function.  The
$(1+z)^4$ bolometric surface brightness dimming of resolved sources is less
significant for such compact sources: the spreading of light rays is
inconsequential if they never spread into more than one psf.  This could lead
to a maximum $(1+z)^2$ relative enhancement of compact sources over resolved,
diffuse sources.  Also, the actual physical distance changes little beyond
redshift $z > 1$, so that a true point source would suffer little $1/r^2$
dimming.  Giant H\ii regions, which have sizes of 0.1 to $1 \kpc$, (Hodge
1993), remain marginally unresolved by HST, and are thus sufficiently compact
for surface brightness dimming to be diminished.  Steidel \etal (1996), and
Cowie \etal (1996) have shown that many sources in the HDF have redshifts $\ga
2.5$; these could produce a 2 magnitude relative enhancement of compact sources
over diffuse sources.

Both of these factors work to enhance the prominence of compact star-forming
regions in a very deep field such as the Hubble Deep Field.

\section{Cataloging the Objects}

We retrieved the HDF version 2 images and version 1 object catalogs from the
Space Telescope Science Institute (Williams \etal 1996).  We found that the
catalog contained spurious faint sources near the margins of the large, bright
(nearby) sources.  This created an overabundance of very close pairs, which
artificially increases the angular correlation function at small scales.  We
therefore created a new catalog, using the {\em daofind} algorithm to identify
objects in deep images formed by averaging the $F814W$ and $F606W$ frames.
This algorithm looks for peaks in the image after bandpass spatial filtering;
the filter is a truncated lowered Gaussian with breadth comparable to the
point-spread function (psf).  {\em Daofind} detected a total of 2817 objects in
the central $71.2'' \times 71.2''$ regions of three wide-field chips.  The
bandpass filtering helps control both random noise and spurious detections in
the wings of very bright sources.  However, it may also wash out object pairs
with separations below about $0.25\arcsec$, and appears to detect subgalaxian
features in some foreground galaxies. To lessen this effect, we masked out
sources brighter than magnitude 23.5 before running our detection algorithms.

\section{Angular Correlation Function}

One may compute the two-point angular correlation function by comparing the
number of data pairs at given angular separation to the number of simulated
random (window) pairs at the same separation.  We thus populated the survey
area with $2 \times 10^4$ random points and computed the distribution of pairs
of data objects with data objects $\langle NN\rangle$, data objects with window
objects $\langle NW\rangle$, and window objects with window objects $\langle
WW\rangle$.  This allowed us to compute (as Hamilton 1993) the angular
correlation for each of the three HDF fields (WFC chips) as
$$w_{est}(\theta) \equiv {{\langle NN\rangle \langle WW \rangle}
\over {\langle NW \rangle^2}} - 1$$
Because the comoving volume of the HDF sample is dominated by $z \ga 1$, most
objects in the catalog are at high redshift.  For reference, we have used two
methods to select higher redshift objects in the catalog.
The first, from Steidel \etal (1996), 
measures spectral curvature between filters to find ``UV dropout''
objects, in which the Lyman-$\alpha$ break has entered the bluest filter.  For
the HDF filter set, this happens at redshift $z\ga 2.5$.  Steidel et al (1996)
find that a useful cut is $F300W-F450W>1.2+[F450W-(F814W+F606W)/2]$.  We
developed a second, similar cut based changing spectral slope, using the filter
selection information provided by STScI (1995), their table 3; this cut prefers
objects with redshift $z_{color} \ga 1.5$.
Any color-based selection criterion ought to place physically
associated objects into the same bin, since such objects presumably
have similar evolutionary histories and hence colors.

We have plotted in figure 1 the angular correlations yielded by the objects
meeting various selection criteria.  In each panel, the lighter curves
represent the correlations from each of the three WFC chips, while the heavier
curve is the mean of those three.  The one-sigma error bars are derived from
comparison of the three values for the individual chips.  The first two rows
gives the correlations for the above-mentioned cuts as applied to our catalog.
The third row, for comparison, contains a catalog from Couch (1996) with the
Steidel \etal (1996) cut applied.

The first (leftmost) column of figure 1 shows the correlations for the higher
color-redshift objects; the second column shows the correlations for the lower
redshift objects; the third column shows the cross-correlations between the
high and low color-redshift objects.  First, we examine the cross-correlation.
If the color cuts effectively sort physically associated objects at different
redshift, there should be little or no correlation between the high and low
cuts.  Indeed, the cross-correlation signal is very low, and consistent with
zero in the first two rows (our catalog).  We thus surmise that these cuts,
applied to our catalog, rarely admit physically associated objects into
different bins.

The high correlations and low cross-correlations in figure 1 tell us that both
moderate and high color-redshift species must exhibit strong correlation on
scales of less than one arcsecond.  This is to be expected at low to moderate
redshift if the cataloging scheme records subgalaxian structure, such as H\ii
regions and bulges.  When one overplots our catalog on the images themselves,
one sees that such structure is detected within galaxies as individual objects.
Some of these galaxies have many detections within them.  These objects,
visibly within single galaxies, likely dominate the correlation signal at low
redshift.  However, at higher redshift, the underlying galaxy may vanish due to
$K$-correction and surface brightness dimming, so that only the compact and
UV-bright star-forming regions are left.  Many object pairs and groups are
visible in the images.  We infer that these pairs and groups are in fact
objects within the same physically associated galaxy, just as the H\ii regions
at low redshift are in the same galaxy.

For this to be so, the correlation scales must correspond to physical scales
which allow detected objects to fit within a single galaxy.  Peebles (1993)
demonstrates that for most cosmologies the angular size of $10\kpc$ galaxies
at $z \sim 2$ is $1.5$-$2.5$ arcseconds ($0.1 < \Omega < 1$, any $\Lambda$),
so that correlations below this scale indicate counting of subgalaxian
objects.  Efstathiou (1991) has shown that projection effects are unlikely to
contaminate correlations, so that small-scale correlation is likely due to
physical association of objects.  The increasing correlation down to
$0.25\arcsec$ implies that many detected objects have sizes $\la
0.25\arcsec$, which is comparable to the expected size ($\sim 0.1''$) of a
$500 \kpc$ H\ii region at $z \ga 1$.  We checked this by computing Petrosian
radii (with $\eta = 2/3$, following the Kron [1995] notation) for the catalog
objects, which showed a broad peak at about $0.15\arcsec$.

If we now compare the high redshift correlations to low redshift correlations
in the first two rows of figure 1, we see that both show substantial
correlation strength on sub-arcsecond scales.  In addition, the high redshift
bin shows a substantially stronger correlation around $0.25\arcsec$.  The
statistical significance of this increase (about one sigma) is debatable,
though a clear excess is visible in all three chips.  
In all three chips, the correlation appears to turn over at half an
arcsecond in the second column, but not in the first.  This excess
correlation on very small scales may be due to the increasing prominence of
very young, compact, and hot star-forming regions at higher redshifts, where
the Wien cut-off begins hiding A type and cooler stars (Fitzpatrick, 1996).
We have confirmed that the correlation persists among a pure
high redshift sample by measuring the correlation functions of both the
high-$z$ candidates and the spectroscopically confirmed high-$z$
sources of Steidel et al (1996).  The peak correlation strengths were
$3.3 \pm 1.2$ and $6.8 \pm 3.5$ respectively.

All of these arguments suggest numerous subgalaxian H\ii regions
in our catalog, both at high $(z \ga 2)$ and lower redshifts.  At low
redshifts, the parent galaxies are visible in the images, but at high redshift,
the parent galaxy is most often invisible, which might lead one to think
na\"\i{}vely that the objects are physically separate, and overcount
``galaxies.''

Finally, we compare correlations derived from our catalog to those derived
from the Couch (1996) catalog, which he created using the {\it SExtractor}
(Bertin 1994) package.  We have found that his catalog does not often overcount
foreground substructure, as does ours.
Correlations in the two catalogs agree at high redshift, 
but disagree significantly in the lower redshift sample.  The smaller
correlation at low redshift and small angular scales in his catalog evinces the
difference between our multiple counting of foreground subgalaxian structure
and his proper counting of one count per one foreground galaxy.  However, the
agreement on the excess sub-arcsecond correlation at high redshift suggests
that both schemes overcount substructure at high redshift.

Couch's SExtractor catalog also shows nonzero correlation at large angular
separations ($1''$--$10''$), absent in our catalog.  This is probably due to
the increased sensitivity of SExtractor's thresholding to faint objects in the
wings of bright objects.  {\it Daofind's} high-pass filtering is less sensitive
to such objects.

As a final test, we made percentile cuts according to angular size (intensity
weighted second moments, $a$ and $b$ from the Couch catalog).  We found
dramatically enhanced sub-arcsecond correlation in small objects.  For the
smallest quartile ($D = \sqrt{ab}< 0.077\arcsec$), we found correlation of
$16\pm 12$ at $0.25\arcsec$; for the smallest 10\% of objects ($D <
0.066\arcsec$), we found correlation of $85\pm 70$ at $0.25\arcsec$ as visible
in the fourth row of figure 1.  This sharp, if uncertain, enhancement
demonstrates that the bulk of the correlation is, in fact, coming from the
smallest objects, in agreement with the hypothesis that the correlation derives
from small, subgalaxian objects.

\section{Number-Magnitude Relation}

We present number-magnitude relations for both unsmoothed and smoothed $F606W$
images in figure 2(a).  The solid histograms represent no smoothing,
the dashed histogram $0.5\arcsec$ smoothing, and the dotted histogram
$1.0\arcsec$ smoothing.  No effort has been made to correct the observed counts
for incompleteness or crowding effects, and the decrease in the counts beyond
$\sim$ 30th magnitude is almost certainly due to incompleteness.  Object fluxes
in $0.16''$, $0.8''$, and $1.6''$ apertures were used to generate the
magnitudes at the three smoothing scales.  We chose aperture photometry for its
robustness in cases where multiple sources overlap on the sky.

The resulting number-magnitude relation rises monotonically to a completeness
limit around AB magnitude $27$ in the $F606W$ filter.  At the faint end, it has
a slope somewhat shallower than that seen in ground-based R band counts, and
substantially shallower than in ground-based U counts (\cf Tyson 1995).  The
number-magnitude slope for the unsmoothed image catalogs steepens at the bright
end, which can be understood as a side effect of using small aperture
photometry.  The smoothed catalogs, on the other hand, show a continued power
law behavior to the brightest objects in the HDF ($AB(F606W) \approx 22$).  The
slope and normalization of the counts are consistent with ground-based counts
for the overlap region ($22 \la R \la 25$) (Tyson 1995).

We have also plotted the $N\propto 1/\mbox{flux}$ line in figure~3.  This line
separates slopes with convergent and divergent total flux at faint magnitudes,
and is a good match to ground-based R band data for $18 \la R \la 25$.  It is
also the number-magnitude slope expected if standard objects are broken into
fragments and then counted as sources (with equal probability for any number of
fragments, up to some large maximum).  Counts from the unsmoothed data match
this slope for $25.5 \la AB(F606W) \la 28$.

Figure 2(b) shows the number-magnitude relation for only the high-color
redshift (using the Steidel \etal [1996] cut).  The results here are not very
different, other than zero-point, from the relation for all sources, which
again suggests that we are seeing a similar population at high and moderate
redshift. 

Comparing the results for $0.5\arcsec$ and $1.0\arcsec$ smoothing to those
for no smoothing, we detect two and four times fewer objects, and the
completeness limit worsens by about two and four magnitudes.  This shows the
deleterious effects of field-crowding in low-resolution fields, where
galaxian images begin to overlap in the poor seeing.  The counts in the
smoothed fields are less likely to overcount substructure, but more likely to
miss faint objects within the seeing discs of bright objects.  Therefore,
care must be taken when comparing ground-based deep counts with counts in the
Hubble Deep Field.

\section{Summary}

We have cataloged objects in the Hubble Deep Field in a way that is less prone
to spurious detections than were previous efforts.  From the catalog, we have
drawn the angular correlation for high and low color-redshift subsets using two
different cuts.  We have found similar correlations down to $0.5\arcsec$
for both subsets.  Since the signal is ostensibly dominated by H\ii regions in
the lower redshift subset, we surmise that that signal is also dominated by
subgalaxian structure in the higher redshift subset.  We have compared our
results to the correlation derived from an independent catalog (Couch 1996)
which appears not to over-count nearby objects. As expected, there is less
correlation in the lower redshift subset of this catalog than in ours.
However, at higher redshift, the correlations of the catalogs agree rather
well, so that both of our catalogs include as distinct objects what are likely
subgalaxian structure at high redshift.  A dramatic increase in sub-arcsecond
correlation occurs in the subset of objects with the smallest angular sizes, in
agreement that the correlated objects are small, subgalaxian objects.

The qualitative difference between deep space- and ground-based
optical data is due to a conspiracy of scales.  Because the
characteristic angular sizes of galaxies at redshifts $1 \la z \la 5$
correspond to the $\sim 1''$ angular resolution of ground-based data,
deep optical counts from the ground will see galaxy-sized objects as
single peaks.  At the higher resolution available from space,
substructure becomes detectable in galaxies at any redshift, and
overcounting becomes a possibility.

We have also computed the magnitude-radius relation, which shows that a large
fraction of the objects have characteristic sizes around $0.15\arcsec$,
corresponding to scale lengths $\sim 1 \kpc$, typical of both high redshift
galaxian scale-lengths and diameters of giant star-forming regions.  The peak
at $0.15\arcsec$ allows several objects to fit into a single galaxy, as one
requires for the subgalaxian structure scenario.

The number-magnitude relations for our catalogs show convergent flux in all
bands with $N \propto 1/\mbox{flux}$ as expected for images broken into
fragments.  This physically reassuring result differs from the na\"\i{}ve
extrapolation of ground-based number-magnitude relations for U, B, and possibly
R bands.  We find that after we smoothed the data, the counts drop dramatically
at the faint end.  This illustrates how seeing reduces the faint counts in
ground-based work, diluting isolated faint objects below detection thresholds
while blurring substructures in brighter galaxies together to form single
peaks.

The statistical tests presented herein suggest that the most distant objects in
the HDF must be some combination of galaxies and star-forming fragments, a
distinction increasingly hard to draw in deep fields.  This supports our
hypothesis that ultraviolet bright and compact star-forming regions contribute
substantially to the flux, and increasingly to the number counts, we receive
from high redshift samples.

\begin{acknowledgements}

WNC is most grateful for the continued support of the Fannie and John Hertz
Foundation, and partial support from NSF grant AST-9529120.  JER and DNS's work
has been supported by NSF grants AST 91-17388, NASA ADP grant 5-2567, and JER's
NSF traineeship DGE-9354937. JER also thanks IPAC for its hospitality.  JPO's
work has been partially supported by NSF grant AST-9424416.  We would also like
to thank Robert H. Lupton, J. Richard Gott, III, Warrick Couch, Sangeeta
Malhotra, and anonymous referees for their very useful communications.  We
thank Jill Knapp for her kind support and encouragement.  Finally, we thank the
HDF team for their hard work and generosity in preparing the data for public
release.

\end{acknowledgements}

\begin{figure}[p]
\plotfiddle{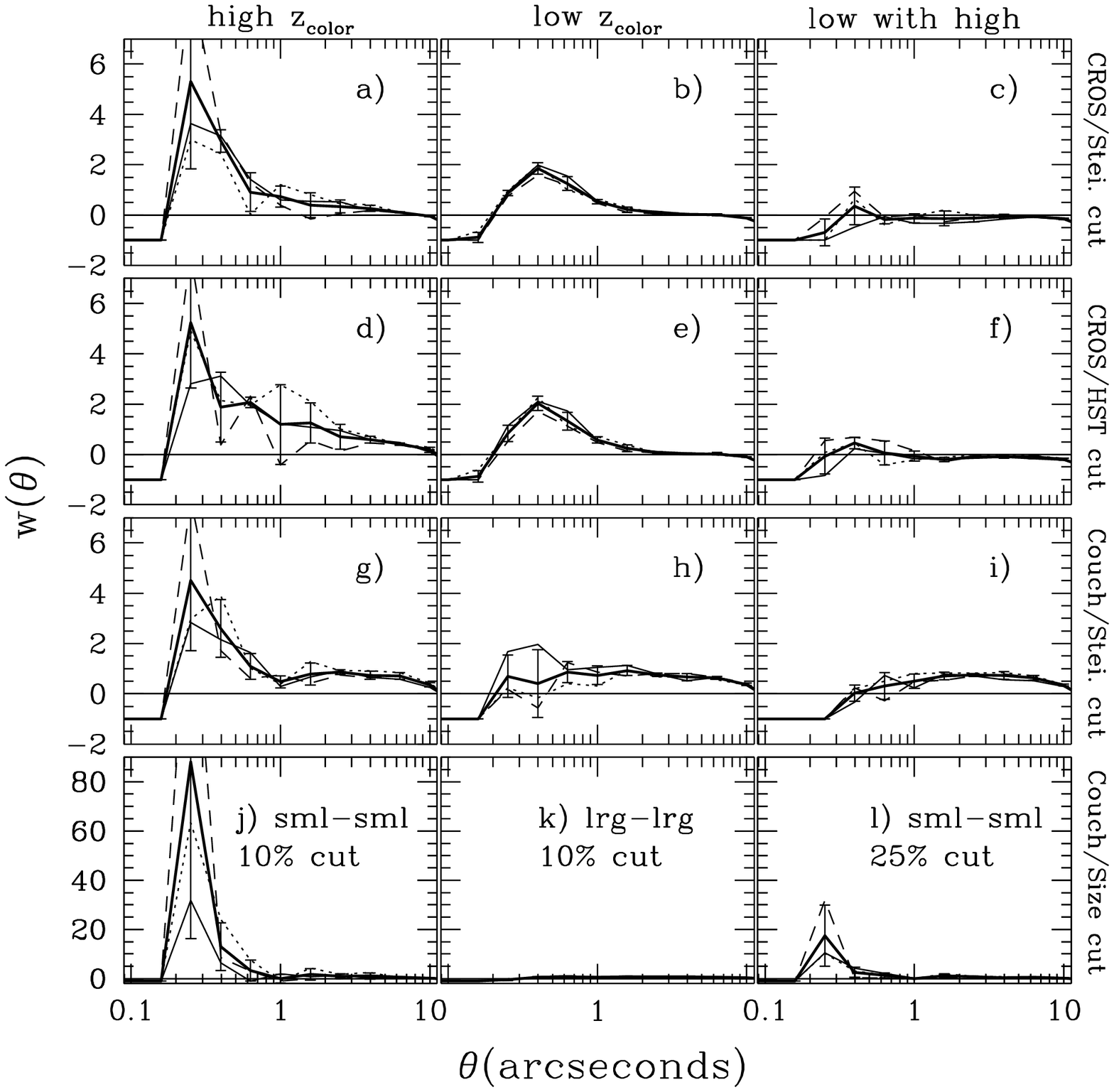}{12cm}{0}{60}{60}{-185}{-90}
\caption{Angular correlation for WFC chips 2, 3, 4 of the Hubble Deep Field.
The lighter curves (solid, dotted, dashed, respectively) are for the individual
chips.  The heavier curve is the mean of those three values.  In the first
three rows, three different catalogs and color-redshift cuts are plotted in
each column.  The top row (a)--(c) is for our catalog with the Steidel
color-redshift cut.  The middle row (d)--(f) is for our catalog with the Space
Telescope cut (see text).  The third row (g)--(i) is for the Couch catalog with
the Steidel cut.  The first column is correlations of high $z_{color}$ objects
with themselves, the second column low $z_{color}$ objects with themselves; the
last column is the correlation of low with high $z_{color}$ objects.  The final
row shows the effect of size cuts in the data. The first column (j) shows
correlations of objects below 10th percentile in diameter.  The second column
(k) shows the correlation of those objects above 10th percentile.  The last
column (l) shows correlations of objects below 25th percentile.}
\end{figure}

\begin{figure}
\plotfiddle{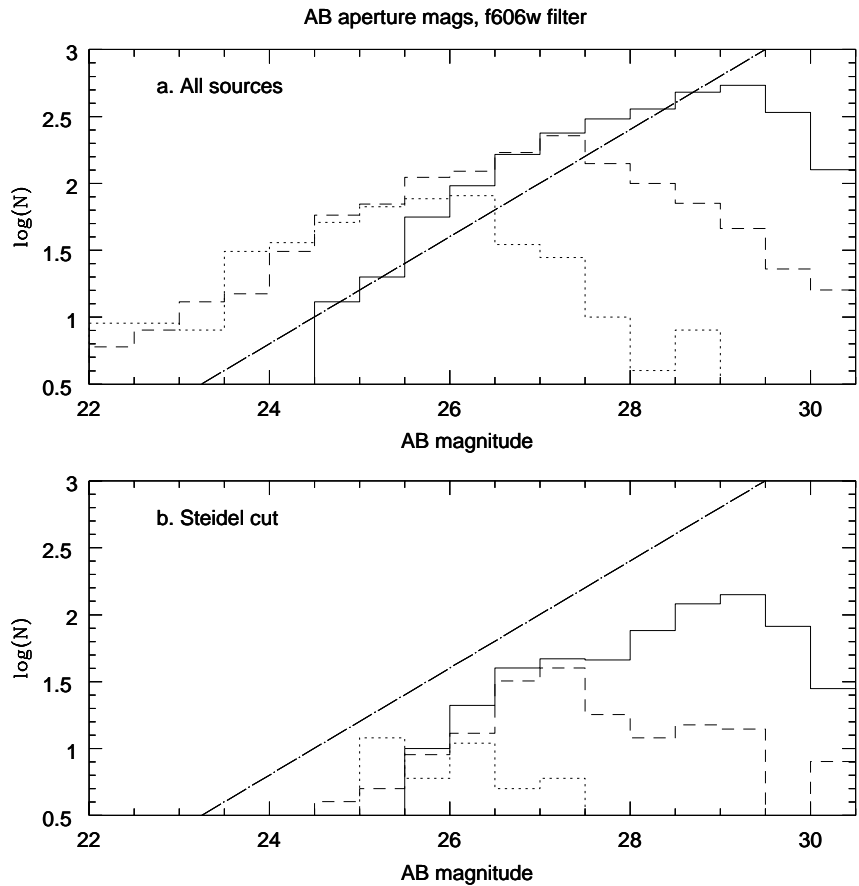}{12cm}{0}{60}{60}{-185}{-90}
\caption{The number-magnitude relations in catalogs derived from unsmoothed
data (solid lines), $0.5\arcsec$ smoothed data (dotted lines), $1.0\arcsec$
smoothed data (dashed lines): (a) the full catalogs, (b) the high color-
redshift subset only.  In (a) the completeness limit decreases by two and four
magnitudes after smoothing, while the number counts decrease by a factor of two
and four, respectively.  The dot-dashed line has slope 0.4.}
\end{figure}

\end{document}